\documentclass[published]{JHEP3} 

\JHEP{00(2010)000}

\JHEPspecialurl{http://jhep.sissa.it/JOURNAL/JHEP3.tar.gz}

\usepackage{epsfig,multicol,bbm}
\usepackage{aas_macros}

\newcommand\fverb{\setbox\fverbbox=\hbox\bgroup\verb}
\newcommand\fverbdo{\egroup\medskip\noindent%
			\fbox{\unhbox\fverbbox}\ }
\newcommand\fverbit{\egroup\item[\fbox{\unhbox\fverbbox}]}
\newbox\fverbbox

\def\beq{\begin{equation}}
\def\eeq{\end{equation}}
\def\bey{\begin{eqnarray}}
\def\eey{\end{eqnarray}}

\def\pc{\, {\rm pc} }

\def\Msun{M_\odot}

\def\a0{$a_0$}

\title{Lopsidedness of cluster galaxies in modified gravity}

\author{Xufen Wu$^1$, HongSheng Zhao$^1$, Benoit Famaey$^2$\\
	
$^1$ SUPA, School of Physics and Astronomy, University of St Andrews, North Haugh, St Andrews, Fife, KY16 9SS, United Kingdom \\$^2$ Observatoire Astronomique, Universit\'e de Strasbourg, CNRS UMR 7550, F-67000 Strasbourg, France\\
	E-mail: \email{xw47@st-andrews.ac.uk}
}

\received{\today} 		
\accepted{\today}		

\preprint{\hepth{9912999}}	

\abstract{We point out an interesting theoretical prediction for elliptical galaxies residing inside galaxy clusters in the framework of modified Newtonian dynamics (MOND), that could be used to test this paradigm. Apart from the central brightest cluster galaxy, other galaxies close enough to the centre experience a strong gravitational influence from the other galaxies of the cluster. This influence manifests itself only as tides in standard Newtonian gravity, meaning that the systematic acceleration of the centre of mass of the galaxy has no consequence.  However, in the context of MOND, a consequence of the breaking of the strong equivalence principle is that the systematic acceleration changes the own self-gravity of the galaxy. We show here that, in this framework, initially axisymmetric elliptical galaxies become lopsided along the external field's direction, and that the centroid of the galaxy, defined by the outer density contours, is shifted by a few hundreds parsecs with respect to the densest point.}


\dedicated{}

\begin{document} 


\section{Introduction}
Modified Newtonian Dynamics \cite{Milgrom1983a,BM1984} is a recipe linking the gravitational field in galaxies with their baryonic content. This recipe makes many successful predictions on galaxy dynamics, such as the baryonic Tully-Fisher relation \cite{McGaugh2005}, and the precise shape of rotation curves of both high and low surface-brightness galaxies in the field \cite{McGaugh_deBlok1998}, as well as those of tidal dwarf galaxies (e.g., Gentile et al. 2007). It could also hold clues to the observed universality of dark and baryonic gravities at the effective dark halo core radius \cite{Donato_etal2009,Gentile_etal2009,Milgrom2009a}. The successes of this recipe in galaxies could be an emergent phenomenon, linked with the complex feedback between baryons and dark matter in the process of galaxy formation \cite{Peebles2009}, but a more radical explanation of these successes is a modification of gravity on galaxy scales. This interpretation requires to postulate additional dark matter at extragalactic scales (e.g., Sanders 2007; Angus et al. 2007; Angus 2009), but let us note that most relativistic versions of such modified gravity theories inevitably include dark fields that seed structure growth, and may play the role of dark matter in galaxy clusters \cite{Ferreira_Starkman2009}. This reduces calculability on large scales, and means that the modified gravity interpretation of the MOND recipe is more easily falsifiable on galaxy scales. Falsifying such an alternative approach to the problems of galactic dynamics is actually needed in order to more firmly anchor the current cold dark matter (CDM) paradigm as the one that should ultimately be confronted with data at the sub-galactic, galactic and cosmological scales.

One of the unique properties of the modified gravity version of MOND in galaxies, compared to its Newtonian CDM counterpart, is that it violates the Strong Equivalence Principle when considering the external gravitational field in which a system is embedded. Hence, the effects of the gravitational environment are important, and fundamental concepts such as the escape speed from a system are fully dependent on the external field. This can lead to extremely interesting and testable predictions. For instance, \cite{Milgrom2009b} analytically predicted in the Solar System an anomaly due to the external field of the Galaxy, in the form of a quadrupole field that may be detected in future measurements of planetary and spacecraft motions. On the other hand, \cite{FBZ2007} and \cite{Wu_etal2007,LMC} studied the effect of the external field from large scale structure on the Milky Way and showed that the MOND predictions for the galactic escape speed were quite realistic, while allowing the Large Magellanic Cloud to be on a bound orbit.

Galaxies inside clusters present a very interesting environment where the external field has a significant influence. These galaxies experience
both their self-gravity and the gravity of the other galaxies of the cluster (except for the galaxy at the centre of the cluster): the latter manifests itself only as tides in standard Newtonian gravity, meaning that the systematic acceleration of the centre of mass of the galaxy has no consequence.  But in the context of MOND, a consequence of the breaking of the strong equivalence principle is that the systematic acceleration changes the own self-gravity of the galaxy.

In this Letter, we emphasize that, in MOND, the combination of the self-gravity with the external acceleration breaks the front-back symmetry of a purely self-gravitational system. As elliptical galaxies are abundant in clusters, they provide an ideal case to test this front-back asymmetry.  Offsets such as those discussed in this letter have also been studied in the recent literature in the context of MOND N-body simulations \cite{phantom_peak}, CDM N-body simulations \cite{LZK2009}, and lensing observations \cite{SFFZ2008}.

\section{Lopsidedness and offset of the centre of gravity}
Let us start with a simple numerical experiment: we place a Plummer sphere into an homogeneous external field typical of a cluster environment, of the order of $\sim 10^{-8} \,$cm~s$^{-2}$, and we solve the modified Poisson equation of MOND, which is a non-linear elliptic partial differential equation~\cite{BM1984}: the boundary conditions are taken as in \cite{LMC}, and the boundary value problem is solved by using the iterative Newton method in spherical coordinates~\cite{CLN2006}. The isodensity and isopotential contours in the free-falling frame are then plotted on Fig.~1, where the external field is applied in the negative $x$-direction: clearly, one sees that the isopotential contours are {\it flattened} and {\it lopsided}, in the sense that they do not exactly correspond to the same radii on the positive and negative $x$-axis. The explanation for this is that, in MOND, the {\it total} gravity (internal+external) enters the dielectric-like $\mu$-function on the left-hand side of the modified Poisson equation \cite{FBZ2007,Wu_etal2007,LMC}: because the external field has a direction, this total gravity is increased by the external field on one side of the galaxy centre along the $x$-axis, and decreased by it on the other side. This effect is maximal when the internal and external gravities are of the same order of magnitude. Clearly, for an originally spherically symmetric distribution, the potential and its derivatives are then still axisymmetric, but the same conclusion holds when applied to an axisymmetric system, and if the direction of the external field does not coincide with the original axis of symmetry, the axial symmetry is broken. We also computed a flattened (axis ratio $0.7$) Plummer model with a diagonally pointing external field.  As expected, we see the twisted and offset potential contours (Fig.2).

\DOUBLEFIGURE[t]{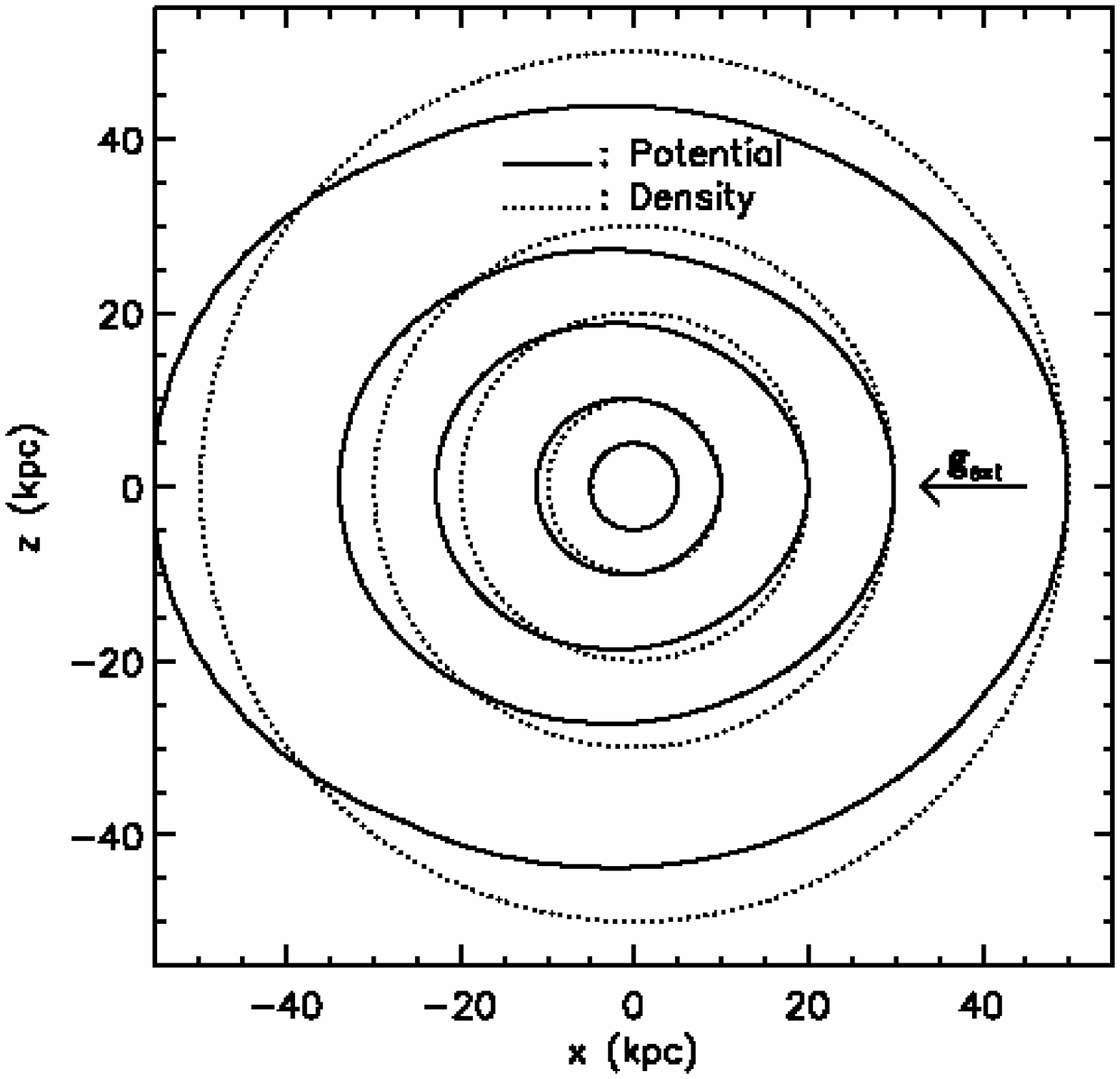, width=.4\textwidth}
{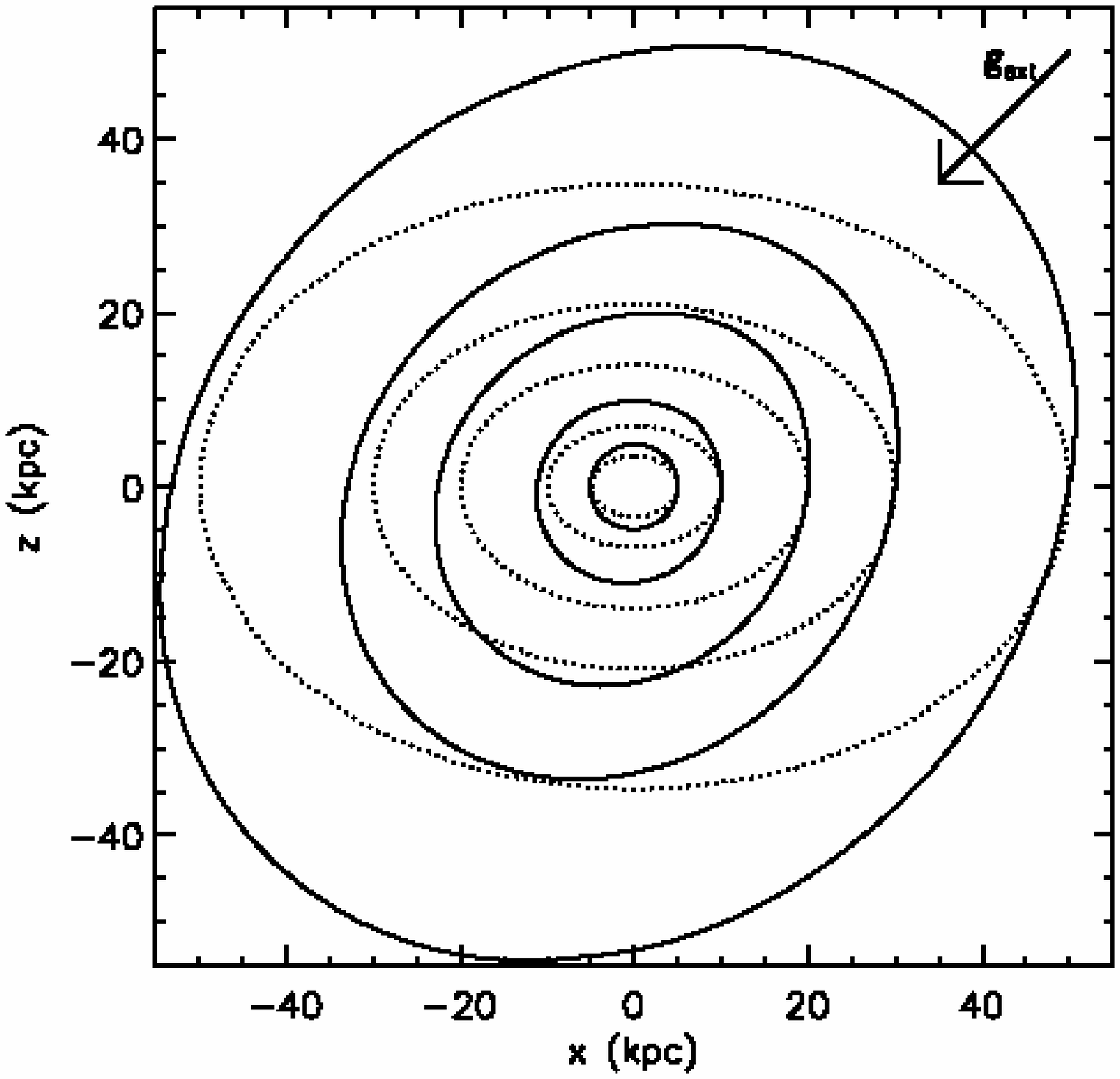, width=.4\textwidth}{Isodensity (dotted) and MOND isopotential (solid) contours of a Plummer sphere with a $5 \times 10^{10} \, {\rm M}_\odot$ mass and a Plummer scale-length of 1~kpc, embedded in an external field of $1.2 \times 10^{-8} \,$cms$^{-2}$ with the direction indicated by the arrow.}{Same as Figure 1 for an axisymmetric Plummer sphere with an axis ratio of $1:1:0.7$.}

An additional interesting effect is the following: because of this lopsidedness of the potential in the outer parts, an additional ``external" field is generated, which acts on the inner parts of the galaxy, akin to the effect in the inner Solar System pointed out by \cite{Milgrom2009b}. This will cause the centre of the galaxy to be slightly shifted with respect to the point where the internal gravity is zero (see also \cite{phantom_peak}). In the Plummer sphere model of Fig.~1, the internal gravity is zero at $x=20 \,$pc. Although this is small, it of course depends on the model parameters, and it means that the initial configuration can never stable and that the photometric centre of the galaxy should shift towards this new centre of gravity, thereby increasing the lopsidedness of the galaxy.
We hereafter investigate whether this is the case with numerical N-body models of realistic elliptical galaxies. Starting from initial conditions built with Schwarzschild modelling, we let the galaxy evolve towards its final state, and check whether it becomes significantly lopsided.

\section{Schwarzschild and N-body models}
As a first step, in order to get our initial conditions, we build static equilibrium models of axisymmetric ellipticals in MOND, with an external field of $1.2 \times 10^{-8} \,$cms$^{-2}$.  The full details of our procedure are given in a companion paper (Wu et al. 2010, submitted). We use the so-called Schwarzschild method \cite{Schwarzschild1979,Zhao1996_bar,triaxial} to reproduce the system self-consistently \footnote{Historically self-consistent galaxy models in Newtonian dynamics mean equilibrium models purely in the gravity of the baryons, without Dark Matter. In MOND this obviously is also the case since no dark matter is supposed to be present. However, the original density reproducing the external field here is not modeled, hence our models are not purely under self-gravity of the model galaxy.} by a linear superposition of orbits in the computed MOND potential of an axisymmetric \cite{Hernquist1990} profile with a total (baryonic) mass ${\rm M}=5\times10^{10} \, {\rm M}_\odot$, and an axis ratio of 0.7. Even when using 27360 orbital building blocks, we still find a significant residue of 4\% between the Schwarzschild model and the input density: this means that a fully self-consistent model {\it cannot} be found for an axisymmetric density model embedded in a strong external field in MOND. We thus expect the system to evolve towards a non-axisymmetric equilibrium configuration.


We then turn this initial Schwarzschild model into a live N-body system by
Monte Carlo sampling of the phases of the orbits. Starting with these initial
conditions, we use the {\it N-MODY} code \cite{NMODY} and let the
system evolve.  Figs.~3, 4, 5 and 6 show the evolution of projected density of the model
after 0, and 90 simulation time (1 simulation time is 1 Keplerian time at the typical scale of 1 kpc, see Eq. (6) in \cite{stability}, $1 T_{simu} = 4.7 Myr$).
One can see clearly an offset of $\sim 200\pc$ of the central densest point from
the centroids defined by the outer projected density contours.  The combination
of orbital anisotropy and the broken axisymmetry makes the evolution quite
complex.  We will report these elsewhere.  By and large this live N-body
simulation confirms our expectation that the axisymmetry is broken in the
presence of an external field, which shows up as a twist of the isophotes and an offset of the density peaks.

\DOUBLEFIGURE[t]{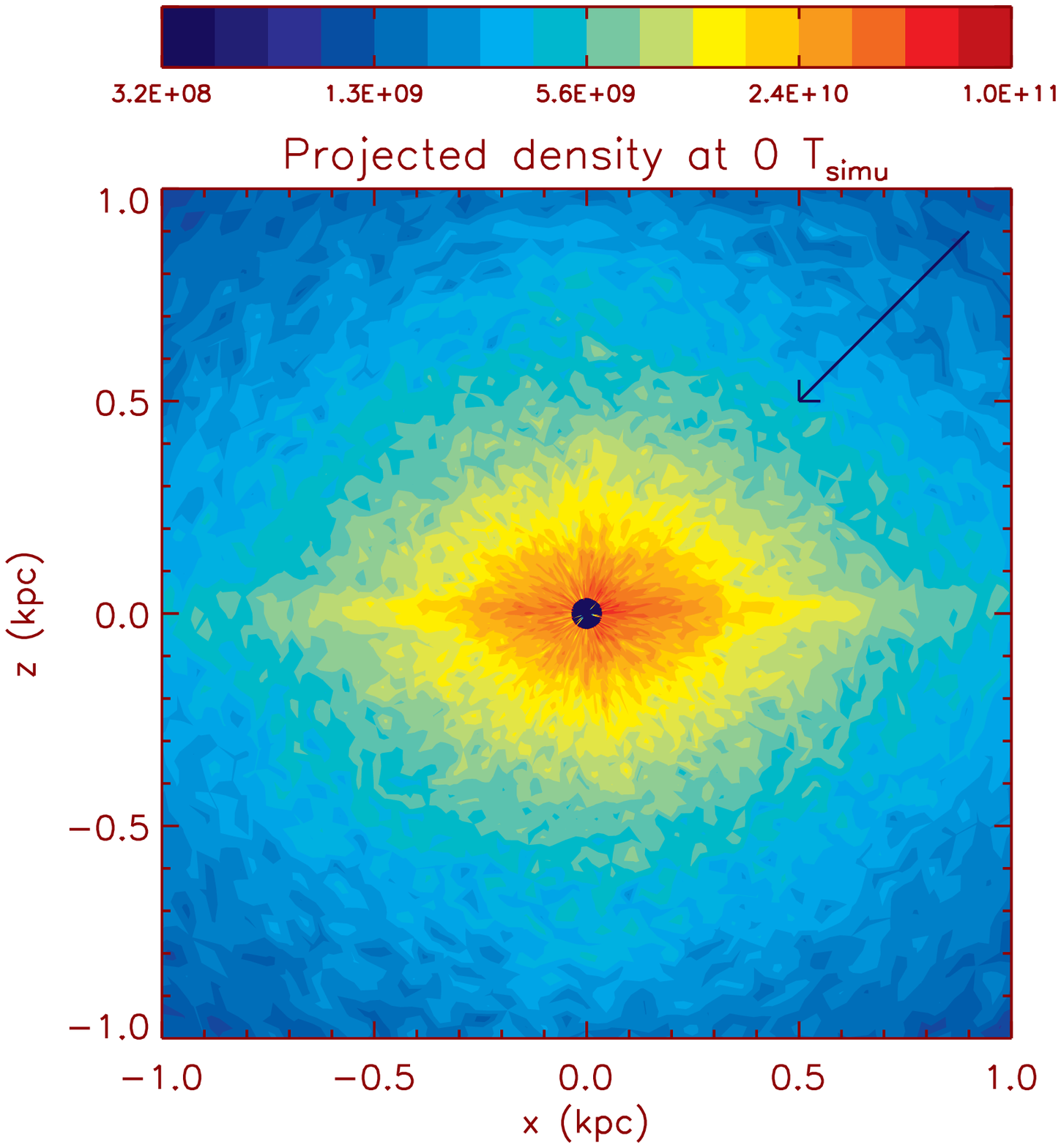, width=.4\textwidth}
{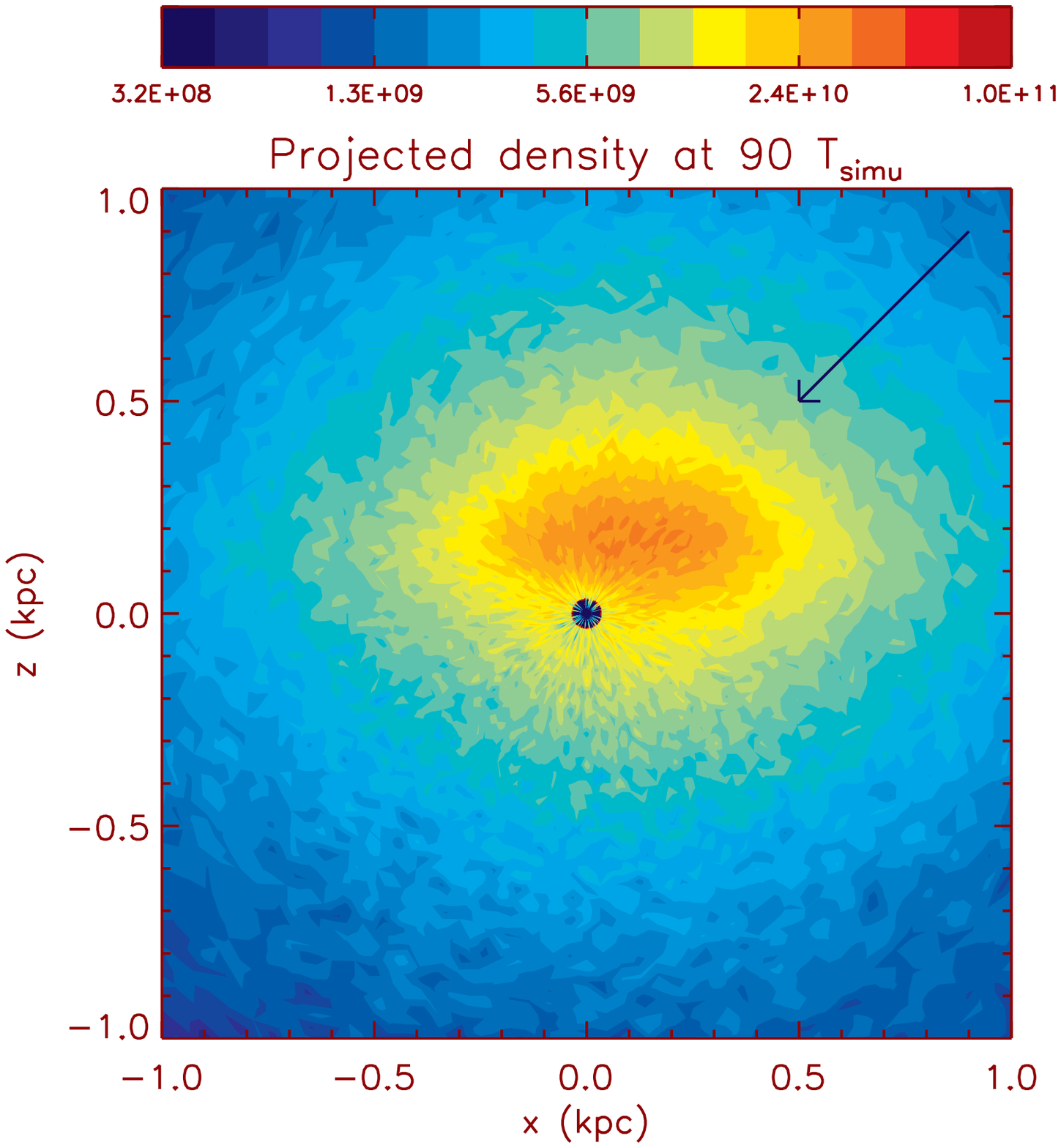, width=.4\textwidth}{Color maps (centroid details on the scale of 1 kpc) of xz-projected density for our initial conditions}{Same as Fig.~3 after t=90 $T_{simu}$. The blue circle in the centre shows the origin point (i.e. centre of mass), roughly corresponding to the centroid defined by the outer surface density contours. It is clearly offset from the central density distribution. The arrows indicate the direction of the external field.}

\DOUBLEFIGURE[t]{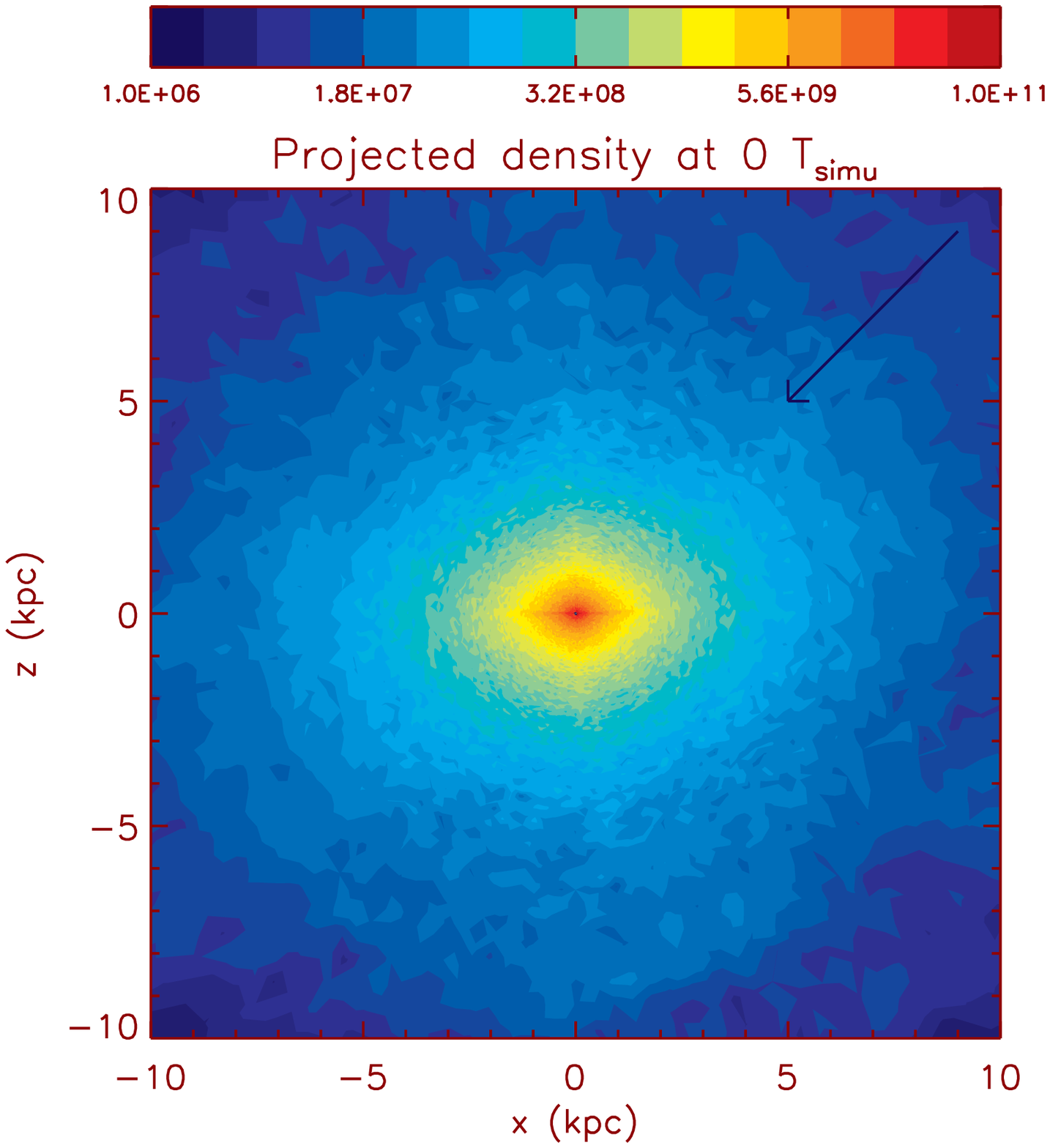, width=.4\textwidth}
{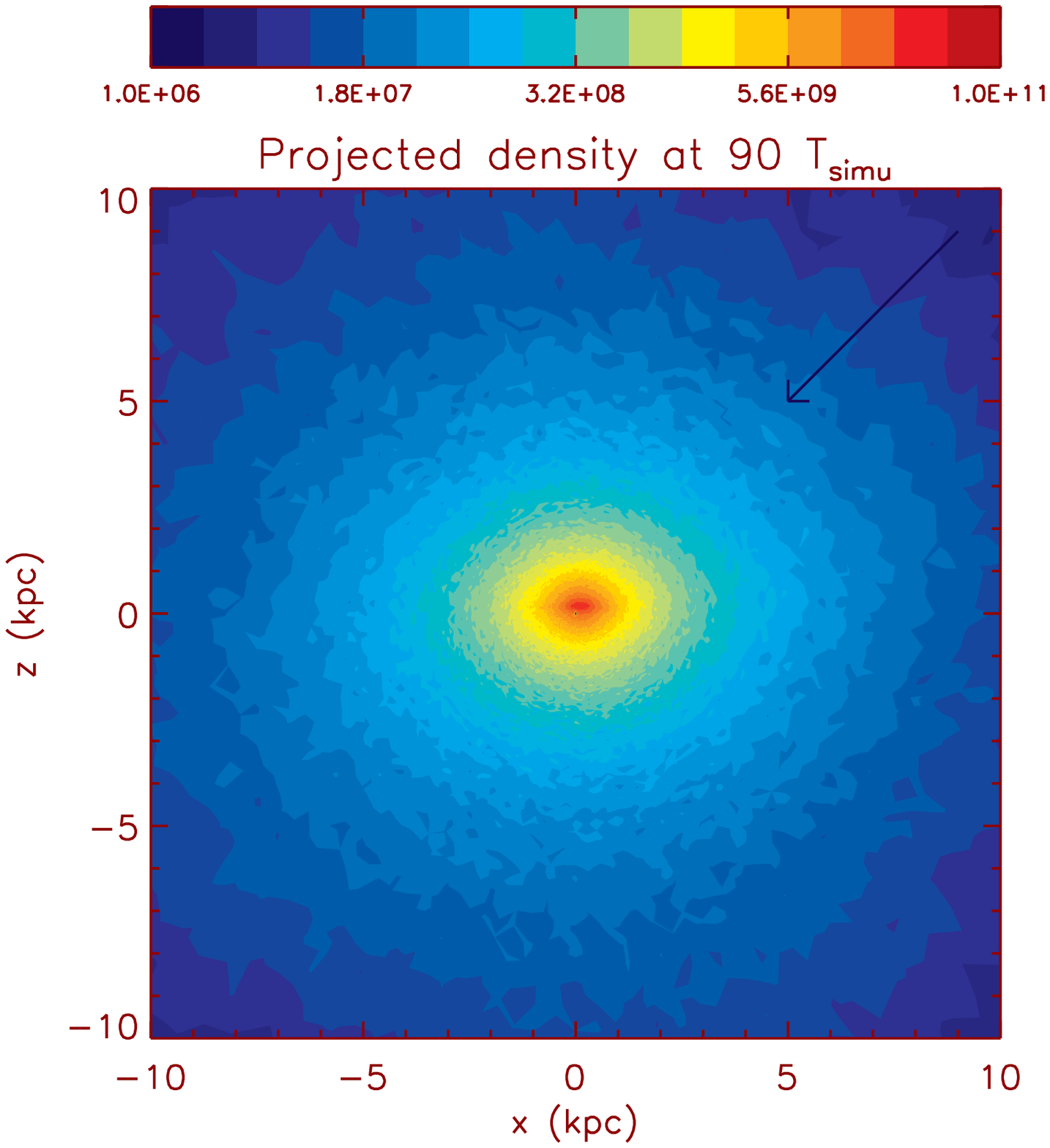, width=.4\textwidth}{Same as Fig.~3 on the scale of 10~kpc}{Same as Fig.~4 on the scale of 10~kpc. This figure shows clearly  the twist of isophotes between the inner and outer density distribution. The arrow indicates the direction of the external field.}



\section{Conclusion}
We estimate that for typical parameters of elliptical galaxies residing in large clusters, a centroid shift of a few hundreds of parsecs is expected in MOND, as compared to the centroid of the outer isodensity contours of the galaxy.  While tides can also cause some
lopsidedness in Newtonian and MONDian gravity, there is a major distinction with the pure MOND effect under scrutiny here. The tidal effects are small inside the tidal radius, and are traditionally neglected when dealing with dense elliptical centres. The fact that these effects can be neglected in our analysis is confirmed by the fact that the average density of our model galaxy inside the saddle point where the external field starts to dominate is larger than the cluster density. The saddle point is estimated to be 15 kpc from the Plummer sphere (total mass $10^{11}\Msun$) centre
(Fig.1  of \cite{phantom_peak}).  Here the average density enclosed is
$\sim 4\times 10^{-3}{\rm M}_\odot {\rm pc}^{-3}$, typically more than one order of magnitudes larger the typical cluster density at that same radius ($1000 \times \rho_{\rm crit} \sim 10^{-4} {\rm M}_\odot {\rm pc}^{-3}$).
Note that \cite{ZT2006} argues that such tidal criteria holds in MOND as well in Newton.

Thus, we conclude that:

\begin{itemize}

\item If real elliptical galaxies in rich clusters show perfectly symmetric light with {\it no} significant offsets between the centroids of the inner and outer contours, the classical version of MOND is likely excluded. This predicted lopsidedness of galaxies inside distant rich clusters should be falsifiable with photometry {\it only}, using, e.g. the VLTI.
To resolve a centroid offset of ~200pc in an elliptical galaxy in
a rich cluster of typical internal gravity $\sim 10^{-8} \,$cms$^{-2}$  at a distance of 160-210 Mpc
(e.g., Abell 1983, Abell 2717, MKW9, \cite{Pointecouteau_etal2005,Sanders2003} would require a
minimum angular resolution of 0.5 arcsec.  The offsets at the centroids of galaxies are between 10 pc and 500 pc: for an elliptical in Coma (100 Mpc far away) this corresponds to 0.02 to 1 arcsec, while for an elliptical in Virgo (20 Mpc far away) it corresponds to 0.1 to 5 arcsec.

\item On the other hand, observing such a lopsidedness would {\it not} consitute a direct falsification of the CDM paradigm. Indeed, CDM simulations predict that isolated CDM halos are typically lopsided due to a lack of relaxation, so this should be the case inside clusters too.
\cite{Maccio_etal2007} studied the shapes of a large sample of ten thousand pure
CDM halos, and found that any offset between the barycentre (CoM) and the density-centre
(potential minimum) is correlated with a massive satellite inside an unrelaxed
halo, i.e., the offset is a measure of unrelaxedness in the CDM context. This is characterized by an offest of order of 1 or 2 percent between the central dark matter density maximum and the centre of mass at the virial radius. Let us note that if this can be translated into a predicted lopsidedness for the light distribution of isolated ellipticals in CDM, this would be a clean test since MOND does not predict such an offset (except at extremely large radii) for isolated galaxies. What is more the centres of some ellipticals are home to strong star formation which could also distort the observed central isophotes. The isophotal distortions predicted in MOND, if actually observed, could thus be due to this. However, for galaxies inside clusters, these more mundane offsets would not be expected to be aligned with the direction of the local external field pointing to the cluster centre. If a statistically relevant preferred direction is found for the lopsidedness of galaxies residing in clusters, it would thus be in support of MOND, although it would not constitute a direct falsification of CDM.
\end{itemize}

Nevertheless, a caveat is that our simple models here were based on a constant external field of $\sim 10^{-8} \,$cm~s$^{-2}$. However, in reality, while galaxies are orbiting inside clusters, the external field acting on them varies with time, both in direction and amplitude. So the next step will be to make the external field vary as a function of time in our simulations, to check how this affects the predicted lopsidedness.

Finally, we also note that the flattened and lopsided potential created by the external field would also be valid in disk galaxies, although most of them are in the field or at the outskirts of galaxy clusters. Still, the flattened MOND potential would create a differential force with a component normal to the disk, hence a specific torque. This could cause differential precession of the disk angular momentum vector, and could lead to the formation of warps, even in a very low external field for isolated galaxies~\cite{CT2009}.

{\bf Acknowledgements:} The work has been performed under the Project HPC-EUROPA (211437), with the support of the European Community - Research Infrastructure Action under the FP8  ``Structuring the European Research Area" Programme. We thank Luca Ciotti, Pasquale Londrillo, Carlo Nipoti for generously sharing their code. XW acknowledges hospitality of the Sterrewacht of Leiden University and of the AIfA of Bonn University. XW acknowledges additional support from HPC-EUROPA (211437). XW is supported by a scholarship of the Scottish Universities Physics Alliance (SUPA).  BF is a CNRS Senior Research Associate. HSZ acknowledges partial support from Dutch NWO fellowship and Royal Society visitorship.

\bibliographystyle{JHEP}
\bibliography{lopsided}

%
%

\end{document}